\documentclass[prl,twocolumn,showpacs,showkeys,amsmath,amssymb]{revtex4}

\usepackage{graphicx}% Include figure files
\usepackage{dcolumn}% Align table columns on decimal point
\usepackage{bm}% bold math
\newcommand{\km}{k_{\rm max}}
\newcommand{\Xla}{X_{\lambda}}
\newcommand{\hXla}{\hat{X}_{\lambda}}
\newcommand{\bfs}{{\bf s}}
\newcommand{\bfz}{{\bf z}}
\newcommand{\Xo}{X^{\ast}}
\newcommand{\E}{\mathbb{E}}

\newcommand{\Kh}{\mathbb{K}}
\newcommand{\etal}{{\it et al.}}

\def\cone{\textrm{$c_d^{(1)}$}}
\def\ctwo{\textrm{$c_d^{(2)}$}}
\def\cthree{\textrm{$c_d^{(3)}$}}

\begin{document}

\preprint{}

\title{Fast Spatial Prediction from Inhomogeneously Sampled Data \\
Based on Generalized Random Fields with Gibbs Energy Functionals
}% Force line breaks with \\
\author{Dionissios T. Hristopulos}
\email{dionisi@mred.tuc.gr }
 \homepage{http://www.mred.tuc.gr/home/hristopoulos/dionisi.html}
\affiliation{Department of Mineral Resources Engineering\\
Technical University of Crete\\Chania 73100, Greece}%
\thanks{}
\author{Samuel Elogne}
\email{elogne@mred.tuc.gr }
\affiliation{Department of Mineral Resources Engineering\\
Technical University of Crete\\Chania 73100, Greece}%

\begin{abstract}
An \textit{explicit} optimal linear spatial predictor is derived.
The spatial correlations are imposed by means of Gibbs energy
functionals with explicit coupling coefficients instead of
covariance matrices. The model inference process is based on
physically identifiable constraints corresponding to distinct terms
of the energy functional. The proposed predictor is compared with
the geostatistical linear optimal filter (kriging) using simulated
data. The agreement between the two methods is excellent. The
proposed framework allows a unified approach to the problems of
parameter inference, spatial prediction and simulation of spatial
random fields.

\end{abstract}
 \maketitle

\section{Introduction}

Spatial prediction of physical variables from samples that are
irregularly distributed in space is a task with applications in many
fields of science and engineering, including subsurface hydrology
\cite{kitan,rubin}, oil reservoir engineering \cite{hohn,sah06},
environmental pollutant mapping and risk assessment \cite{christ},
mining exploration and reserves estimation \cite{goov},
environmental health studies \cite{ch98}, image analysis
\cite{winkler} and neuroscience \cite{leow}. Physical quantities of
economical and environmental interest include mineral grades,
concentrations of environmental pollutants, soil and rock
permeability and flow fields in oil reservoirs. Modeling the
variability of such processes is based on the theory of spatial
random fields (SRFs) \cite{yaglom}. Knowledge of spatial
correlations in SRFs enables (i) generating predictive isolevel maps
(ii) estimating prediction uncertainty and (iii) developing
simulations that reconstruct probable scenarios conditioned on the
data. The classical approach is based on Gaussian SRF's (GSRF's) and
various generalizations for non-Gaussian distributions
\cite{lantu,wack}. For GSRF's the spatial structure is determined
from the covariance matrix, which is estimated from the available
sample(s).

Let $\Omega \in {\mathbb R}^{d}$ denote the area of interest, and
$|\Omega|$ its volume. An SRF state (realization) in $\Omega$ can be
decomposed into a {\it deterministic trend} $m_{\rm x} ({\bf s})$, a
{\it correlated fluctuation} ${X}_{\lambda}({\bf s})$, and an
independent random noise term, $ \epsilon({\bf s}) $, i.e.,  $
X({\bf s})=m_{\rm x} ({\bf s})+{X}_{\lambda}({\bf s})+\epsilon({\bf
s}).$ The trend represents large-scale variations obtained in
principle by ensemble averaging, i.e. $m_{\rm x} ({\bf s})=E[X({\bf
s})]$. In practice, the trend is often determined from a single
available realization. The fluctuation represents `fast variations'
describing fine structure above the resolution limit $\lambda$. The
random noise represents non-resolved inherent variability, purely
random additive noise, or non-systematic measurement errors. The
fluctuation typically is a {\it second-order stationary SRF}, or an
{\it intrinsic SRF} with second-order stationary increments
\cite{yaglom}. The  {\it residual SRF} after trend removal is a
zero-mean fluctuation: $ X^{*}({\bf s})=X_{\lambda}({\bf
s})+\epsilon({\bf s}).$ The inference process focuses on determining
a model for $X_{\lambda}({\bf s})$ from a set of possibly noisy
observations $ X^{*}({\bf s})$.

In statistical physics the {\em probability density function (pdf)}
of any fluctuation field $X({\bf s})$ governed by an energy
functional $H[X({\bf s})]$  is expressed as $f_{\rm x} [X({\bf s})]
= Z^{- 1} \exp \left\{ { - H[X ({\bf s})]} \right\},$ where $ Z $ is
the partition function. In classical geostatistics, the Gaussian
joint pdf for a set of fluctuations ${\bf X}=\{X({\bf s}_{i}),
i=1\ldots,N\}$ is expressed in terms of $H[{\bf X}] = \frac{1}{2}
X({\bf s}_i) \, [C_{x}]^{-1}_{ij} \, X({\bf s}_{j})$, where
$[C_{x}]^{- 1}_{ij}$ is the inverse covariance matrix, and summation
is implied over repeated indices. Instead, Spartan Spatial Random
Fields (SSRF's) \cite{dth03} model spatial correlations in terms of
`interactions'. This change in viewpoint has important consequences
for model inference and spatial prediction.

\section{The FGC model}
In \cite{dth03} the fluctuation-gradient-curvature (FGC) SSRF model
is defined and its properties investigated. The continuum FGC model
involves the following $H[{\bf X}]$:

\begin{equation}
\label{fgc} H_{\rm fgc} [X_\lambda  ] = \frac{1}{{2\eta _0 \xi ^d
}}\int {d{\bf s}}   \left[  S_0(\bfs) + \, \eta _1 \,\xi ^2 \,
S_1(\bfs) + \xi ^4 \, S_2(\bfs) \right],
\end{equation}
where $ S_0(\bfs)= \left[ {X_\lambda ({\bf s})} \right]^2$,
  $S_1(\bfs) =\left[ {\nabla X_\lambda ({\bf s})} \right]^2$, and
    $S_2(\bfs) =\left[{\nabla ^2 X_\lambda ({\bf s})} \right]^2 $.
The model is characterized by four parameters: the scale coefficient
$\eta_0$, the covariance-shape coefficient $\eta_1$, the
characteristic length $\xi$, and the cutoff wavevector $\km$. {\it
Bochner's permissibility theorem} \cite{christ} for the positive
definiteness of the covariance function requires $\eta _1> -2$ if
$\km \rightarrow \infty$. In statistical physics terminology,
$\ell_{1} =\eta_1 \,\xi^{2-d}/(2\,\eta_0)$ and
$\ell_{2}=\xi^{2-d}/(2\,\eta_0)$ represent the coupling strengths of
the gradient and curvature terms. A coarse-graining kernel is used
to cut off the fluctuations at $\km $
 \cite{dth03,dthel06}, leading to band-limited
covariance spectral density. If $k_{max}$ is finite, the field's
configurations are almost surely differentiable  \cite{dthel06}. If
$\km$ is infinite, generalized gradient and constraints should be
used. The coarse-graining kernel implies that the SRF $\Xla$ is a
{\it generalized SRF} \cite{yaglom}.

A moment-based method for parameter estimation was proposed and
validated with simulated data \cite{dth03}. The inference process is
based on matching ensemble constraints $\E[S_j(\bfs)]$ with their
sample counterparts, denoted by $\overline{\mathcal{S}_j(\bfs)}$,
for $j=0,1,2$. The procedure is extended in \cite{eldth06}.

Assume $S_{\rm m}=\{{\bf s}_1, \ldots {\bf s}_N \}$ is a set of
sampling points on an irregular grid and $X^{*}(S_{\rm
m})=\{X^{*}_{1},\ldots,X^{*}_{N}\}$ is the respective vector of
measurement. On an irregular grid, the translation symmetry of the
lattice is lost. The continuum FGC functional is then a more
suitable model. For practical purposed, a tractable approximation of
the continuum model is needed.

In \cite{eldth06}, approximations for the sample averages of
$S_1(\bfs)$ and $S_2(\bfs)$ are formulated in terms of kernel
averages of the data values. In the following we use the notation:
$c_d^{(0)}=4d(d+2),$ $\cone=d,$ $\ctwo=4d^2$ and $\cthree=2d(d-1),$
$\left\langle A_{i,j} \right\rangle_{h} \equiv \frac{\sum_{i \neq j}
A_{i,j} \Kh_{h}(\bfs_{i}-\bfs_{j})}{\sum_{i \neq j}
\Kh_{h}(\bfs_{i}-\bfs_{j})}$, where the summation is over both
indices $(i,j=1,\ldots,N)$ denotes the average of the quantity
$A_{i,j}$, weighted by the kernel $\Kh_{h}({\bf r})$ with {\it
bandwidth} parameter $h$. The pair distance is denoted by $s_{i,j}
=\| {\bf s}_{i}-{\bf s}_{j}\|$, where $\|{\bf r}\|$ is the Euclidean
norm of the distance vector ${\bf r},$ and the field increment by
$\Xo_{i,j} \equiv \Xo(\bfs_{i}) - \Xo(\bfs_{j})$. Then, the
\textit{generalized gradient constraint} is given by
\begin{equation}
\label{eq:barS1irr} \overline{\mathcal{S}_1(\bfs)}=
\frac{\cone}{a_{1}^{2}} \,
 \left\langle  \left( \Xo_{i,j} \right)^2
 \right\rangle_{h_1}
 \end{equation}
 where $a_1^{2} = \left\langle s_{i,j}^2 \right\rangle_{h_1} $.
   Sensible estimates of the spacing
$a_1$ should account for the grid topology. E.g., let
$\mathfrak{B}_0$ be the set of near-neighbor vectors of all the
points in $S_{\rm m}$. If $\mathfrak{B}_0$ contains $N_0$ vectors
and $\Delta_i$ denote the lengths of the vectors in
$\mathfrak{B}_0$, then $ \hat{a}_1^{d}= \frac{1}{N_0} \sum_{i
=1}^{N_0} \Delta_i^d $. Similarly, if $a_2^{4} =\left\langle
s_{i,j}^4 \right\rangle_{h_2}$, $h_3=\sqrt{2}\,h_2$,  $h_4=2\,h_2,$
the \textit{generalized curvature constraint}
$\overline{\mathcal{S}_2(\bfs)}$ is given by

%\begin{widetext}
\begin{eqnarray}
\label{eq:barS2irr} \overline{\mathcal{S}_2(\bfs)}& =
&\frac{1}{a_{2}^{4}} \left\{ \ctwo \mu_1  \,
 \left\langle  \left( \Xo_{i,j} \right)^2
 \right\rangle_{h_2}-\cthree \mu_2
 \left\langle  \left( \Xo_{i,j} \right)^2
 \right\rangle_{h_3} \right.
 \nonumber \\
 &- & \left. \cone\left\langle  \left( \Xo_{i,j} \right)^2
 \right\rangle_{h_4} \right\},
 \end{eqnarray}
 where $\mu_1$ and $\mu_2$ are $1+o(\epsilon)$ constants that depend on the sampling
 network
 topology.  The $\mu_1$,  $\mu_2$ are defined so as to satisfy asymptotic bias
and consistency properties \cite{eldth06}. They introduce explicitly
in the problem of model inference four parameters linked to the
topology of the sampling network: the spacings $a_1$ and $a_2$ that
replace the lattice constant, and the bandwidths $h_1$ and $h_2$
that determine the range of influence of the averaging kernel. The
latter are determined from the \textit{consistency principle}
$a_p^{2p} =\left\langle s_{i,j}^{2p} \right\rangle_{h_p},$ where
$p=1,2.$

\section{Spatial Prediction}
\label{spatest} Let $Z_{\rm p}=\{{\bf z}_1, \ldots {\bf z}_K\}$ be a
set of prediction points, disjoint from $S_{\rm m}$, $V_{l}=S_{\rm
m} \cup \{{\bf z}_l\}$, and $V=Z_{\rm p} \cup S_{\rm m}$. The
predictions will be denoted
 by $\{ \hat{X}_{\lambda}({\bf z}_{l}), \, l=1,\ldots, K \}$ and
 the respective prediction vector by $\hat{\Xla}(Z_{\rm
p})$. The increments corresponding to  $V_{l}$ will be denoted by
$\alpha_{p}(V_{l})$, $p=1,2$. Typically, single-point prediction is
applied sequentially over all points in $Z_{\rm p}$. Multiple-point
prediction is possible in the SSRF framework, but this letter
focuses on single-point prediction.

\subsection{Optimal Linear Prediction}
\label{kriging} In geostatistics, spatial prediction is based on the
Best Linear Unbiased Estimator (BLUE), commonly known as Kriging
\cite{kitan,wack}. Different variants of kriging exist, depending on
the hypotheses about the normality of the data and the behavior of
the mean. These methods are generalizations of the linear minimum
mean square error (LMMSE) estimators, also known as Wiener filters
\cite{ra05}.  \textit{Ordinary kriging} (OK) is the most common
variety. It is applied to normally distributed data, with an unknown
mean that can be considered as locally constant. A {\it
single-point} prediction is obtained as a superposition
$\hat{X}(\bfz_l)= \sum_{j=1}^{M} \lambda_j \, X(\bfs_j)$, where
$\bfs_j$ are all the points inside a local search neighborhood,
$B({\bf s}_{l})$, around $\bfz_l$. The \textit{prediction error} is
defined as $\varepsilon(\bfs_l)=\hat{X}(\bfz_l)-X(\bfz_l)$. The
optimal linear coefficients should minimize the mean square error
conditional on the zero-bias constraint $\sum_{j=1}^{M} \lambda_j =
1$, i.e., the expression $\E\left[\varepsilon^{2}(\bfs_l) \right] +
\mu \left( {\sum}_{j=1}^{M} \lambda_j - 1 \right),$ where $\mu $ is
a Lagrange coefficient. This leads to the linear system $C_{\rm
X}(\bfs_i-\bfs_j) \, \lambda_j +\mu = C_{\rm X}(\bfs_i-\bfz_l)$,
$\forall \, i=1,\ldots,M$ and $\sum_{j=1}^{M} \lambda_j = 1$, where
$C_{\rm X}({\bf r})$ is the centered covariance function. OK is an
exact interpolator, meaning that $\{\hat{X}(\bfs_{i}) = X^{*}
(\bfs_{i}), \forall {\bfs}_{i} \in S_{\rm m}\} $. Exactitude is not
always desirable, since it ignores measurement errors and
over-constrains the predictions. For a fixed-size search
neighborhood, the numerical complexity of an $M$-point OK prediction
is $O(K\, M^{3})$.

\subsection{Spatial Prediction based on FGC Functional}
\label{fgc-prediction} Once the parameters of the FGC model have
been determined from the data, prediction of the SRF at $\bfz_l$ is
possible by means of at least two different approaches. First, the
corresponding covariance function is determined and spatial
predictions are then obtained using the OK predictor.  In this case,
the only difference introduced by the SSRF functional is the
covariance estimator.

Here we propose a different predictor, obtained by  maximizing the
conditional probability density $f_{\rm X} \,[\,\Xla (\bfz_l )\, |
\,\Xla(S_{\rm m}) \,]$. Since $f_{\rm X} \,[\,\Xla (\bfz_l )\, |
\,\Xla(S_{\rm m}) \,]=f_{\rm X} \,[\, \Xla(V_{l}) \,]\, / f_{\rm X}
\,[\Xla(S_{\rm m})]$, the prediction is obtained by maximizing
$f_{\rm X} \,[\Xla (V_{l})]$. In principle, this requires solving
the equation $\delta H_{\rm fgc}/\delta \Xla(\bfz_l)=0$, where
$\frac{\delta [.]}{\delta\Xla(\bfz_l)}$ is the variational
derivative of the functional given by Eq.~(\ref{fgc}) with respect
to $\Xla(\bfz_{l})$. In practice, $H_{\rm fgc}$ is replaced by a
discretized estimator, $\hat{H}_{\rm fgc}$. Since $\hat{H}_{\rm
fgc}$ is a bilinear functional, the prediction follows from the
solution of the linear equation:

\begin{equation}
\label{eq:pred-H1} \left. \frac{\partial \hat{H}_{\rm
fgc}\,[\,\Xla(V_{l}) ]} {\partial \Xla(\bfz_l )}
\right|_{\hXla(\bfz_l) } =0.
\end{equation}
$\hat{H}_{\rm fgc}$ is obtained by means of the estimators $\{
\overline{\mathcal{S}_j(\bfs)}, j=0,1,2 \},$ using the
\textit{ergodic hypothesis}
 $ \int d\bfs \, S_{j}(\bfs) \approx |\Omega| \,
\overline{\mathcal{S}_{j}(\bfs)},$ which leads to:
\begin{equation}
\label{eq:fgc-app} \hat{H}_{\rm fgc} [\,\Xla(V_{l})  ] =
\frac{|\Omega|}{{2\eta _0 \xi ^d }}   \left[
\overline{\mathcal{S}_0(\bfs)} + \, \eta _1 \,\xi ^2 \,
\overline{\mathcal{S}_1(\bfs)} + \xi ^4 \,
\overline{\mathcal{S}_2(\bfs)} \right],
\end{equation}
where the spatial averages involve the sampling points and the
prediction point as well. In light of (\ref{eq:barS1irr}),
(\ref{eq:barS2irr}), and (\ref{eq:fgc-app}), equation
(\ref{eq:pred-H1}) leads to the following linear predictor

%\begin{widetext}
\begin{equation}
\label{pred-1} \hat{\Xla}(\bfz_l )=\frac{\sum_{i=1}^{4}
q_{i}\,c_i(V_l) \, \langle \Xo \rangle_{h_i}^{l}} {1+\sum_{i=1}^{4}
q_{i}\,c_i(V_l)},
\end{equation}
%\end{widetext}
where $q_{1}=q_{2}=1$, $q_{3}=q_{4}=-1 $, $\langle \Xo
\rangle_{h}^{l}$ is the kernel average of the sample, centered at
the prediction point \begin{equation} \label{kern-av} \langle \Xo
\rangle_{h}^{l}= \frac{\sum_{i}\Kh_{h}(\bfs_{l}-\bfs_{i}) \, \Xo_i}
{\sum_{i}\Kh_{h}(\bfs_{l}-\bfs_{i})}, \end{equation} \noindent and
the linear coefficients $\{c_{i}, i=1,2,3,4 \}$ are given by
$c_i(V_l)   =    b_i(V_l) \, (N+1) \, g_{h_i}(V_{l}),$ $ b_1(V_l)
=\cone \eta_1 \check{\xi}_1^{2},$
 $b_2(V_l)  =  \ctwo\,\mu_1(V_l)\, \check{\xi}_2^{4},$
 $b_3(V_l)  =  \cthree\,\mu_2(V_l)\, \check{\xi}_2^{4},$
  $b_4(V_l)  =  \cone\, \check{\xi}_2^{4},$ $\check{\xi}_p=\xi/a_p(V_l)$
  and
 $$ g_{h}(V_{l})  =
\frac{\sum_{i}\Kh_{h}(\bfs_{l}-\bfs_{i}) }{\sum_{j> i}
\Kh_{h}(\bfs_{i}-\bfs_{j})+\sum_{i}\Kh_{h}(\bfs_{i}-\bfs_{l})}.$$
 The summation in $g_{h}(V_{l})$ extends over
all the $N(N-1)/2$ non-identical and non-repeating pairs of sampling
points. Defining the linear weights
$$\lambda_{i}(V_l)=\frac{(-1)^{\delta_{i>2}}\,
c_i(V_{l})}{1 + c_1(V_l) +c_2(V_l) - c_3(V_l)-c_4(V_l)},$$
 the prediction is expressed as
\begin{equation}
\label{pred-2} \hat{\Xla}(\bfz_l )= {\sum}_{p=1}^{4}\,
\lambda_{p}(V_l) \,\langle \, \Xo \, \rangle_{h_p}^{l}.
\end{equation}

\subsection{Properties of the FGC Mode Predictor}
The present formulation of the FGC mode predictor (FGC-MP) is closer
to simple kriging than OK, since the mean is assumed to be known.
However, unlike simple kriging the mean does not have to be
constant, provided that it changes slowly so that the energy
contributions due to the square gradient and curvature of the mean
can be ignored compared to the fluctuations. In this respect, the
predictor resembles OK, which allows for slow (but unknown)
variation of the mean. Predictions of the FGC-MP  are independent of
$\eta_0$, while the prediction variance is linearly proportional to
$\eta_0$.

The FGC-MP is linear and unbiased. Since the joint FGC pdf is
Gaussian, the mode estimate is equivalent to the minimum mean square
estimate. Thus, the FGC-MP is an optimal linear predictor. The main
differences with OK result from the use of the energy functional in
the FGC SSRF: (1) The FGC-MP is not an exact interpolator, because
it does not use the data at the prediction point. (2) The
single-point FGC-MP provides an explicit expression for the
prediction, while kriging requires solving a linear system. (3) The
FGC-MP does not require specifying a search neighborhood around the
prediction point; in kriging definition of a search neighborhood
requires an iterative procedure based on cross-validation of the
predictions with the data. (4) The FGC-MP incorporates two sets of
parameters: the first set determines the spatial dependence of the
SRF, while the second set depends on the topology of the sampling
network. The influence of the sampling topology is not explicitly
accounted for in kriging. (5) The uncertainty estimate involves the
SSRF covariance function, for which there are no explicit solutions
in $d=2$, unlike $d=1,3$ \cite{dthel06}. Obtaining the covariance in
$d=2$ requires performing numerically a univariate (for isotropic
dependence) integration of the spectral density. (6) Regarding
multiple-point estimates the FGC-MP has a numerical complexity
$O(K^3)$, derived from solving a linear system of $K$ coupled
equations at the prediction points, while the numerical complexity
of kriging is $O(K\,M^{3})$.

\section{Prediction using Simulated Samples}
\label{sim-pred} At $400$ randomly distributed points on a square
domain of length $L=100$ we generate $100$ independent ``samples''.
These represent realizations of a Gaussian random field with $m_{\rm
x}=50,$ and an  exponential covariance function
 $C_{\rm x}({\bf r})= \sigma_{\rm x}^{2} \, \exp(-\|{\bf r}\|/b_e)$,
 where $\sigma_{\rm x}=10,$ and $b_e=10.$
 The Cholesky LU decomposition method is used for the
 simulations. We partition the $400$ points into a training set
  $S_{\rm m}$ of $100$ randomly selected points, and a prediction
  set, $Z_{\rm p}$,  including the remaining points.
We use the first set to determine the optimal SSRF parameters, and
then predict the values of the field at the locations of the
prediction set. The triangular kernel is used in the FGC-predictor
mode.  Predictions are also generated using the Ordinary Kriging
method.

The performance of the predictors is evaluated using  the bias, the
mean absolute error (mae), the root mean square error (rmse), the
mean absolute relative error (mare), the root mean square relative
error (rmsre) and the linear correlation coefficient ($R^{2}$). The
means are calculated with respect to the values at the $300$
prediction points. Statistics of these quantities over the $100$
samples are shown in Tables~\ref{predtab1} and \ref{predtab2}. The
kriging predictor is applied with the {\it a priori} parameters of
the exponential covariance (instead of the inferred covariance model
from the data). This choice aims at testing the FGC-Mode Predictor
against the ``true'' model. The results show that the two predictors
perform very similarly.
\begin{table}
\caption{\label{predtab1} Statistics of OK performance. }
\begin{ruledtabular}
%\begin{center}
\begin{tabular}{lcccc}
  &  {\bf Minimum} & {\bf Maximum} &
  {\bf Mean} & {\bf Median}
  \\ \hline
 { bias} &  $-1.90$ & 2.10 & $-0.04$ & $-0.14$
  \\
 { mae } &  5.00 & 6.95 & 6.03 & 6.11
  \\
 { rmse } & 6.28 & 8.98 & 7.76 & 7.81
  \\
 { mare } &  0.10 & 0.15 & 0.12 & 0.12
  \\
 { rmsre } &  0.13 &0.25& 0.17 & 0.17
  \\
 { R$^{2}$ } &  0.38 & 0.76 & 0.62 & 0.63
   \\
\end{tabular}
%\end{center}
\end{ruledtabular}
\end{table}
%===============================================
\begin{table}
\caption{\label{predtab2} Statistics of FGC-Mode performance. }
\begin{ruledtabular}
\begin{tabular}{lcccc}
  &  {\bf Minimum} & {\bf Maximum} &
  {\bf Mean} & {\bf Median}
  \\ \hline
 { bias} &  $-1.58$ & 2.40 & $-$0.03 & $-$0.00
  \\
 { mae} &  5.07 & 7.07 & 6.13 &6.10
  \\
 { rmse} & 6.40 & 9.04 & 7.73 & 7.73
  \\
 { mare} &  0.10 & 0.16 & 0.13 & 0.13
  \\
 { rmsre} &  0.13 &0.27& 0.18 & 0.17
  \\
 { R$^{2}$ } &  0.37 & 0.76 & 0.60 & 0.61
   \\
\end{tabular}
\end{ruledtabular}
\end{table}
\section{Conclusions}
A fast linear optimal  predictor, with applications in the analysis
of spatial data, is proposed. The predictor is based on generalized
random fields which represent the spatial dependence in terms of
interactions. An explicit expression for single-point prediction is
obtained. The reduced numerical complexity of the FGC-Mode predictor
may promote the use of cross-validation procedures for model
parameter inference, instead of the commonly used parametric
methods.  The SRF representation, which is based by construction on
an objective function, provides a unified framework for model
parameter estimation, spatial prediction and constrained (respecting
the data) simulation. This is in contrast with classical approaches
that require the introduction of {\it ad hoc} objective
functions~\cite{goov, sah06} for simulations (e.g., by means of
simulated annealing.) Expressions for the prediction uncertainty and
a linear system  for multiple-point prediction have also been
derived and will be reported elsewhere \cite{dthel07}. The
multiple-point predictor accounts for interactions between the
prediction points that may lower the total ``energy''. Such
interactions are missed in single-point prediction. Finally, the FGC
focuses on short-range interactions, but long-range dependence can
be incorporated in the SSRF framework with suitable modifications of
the energy functional.

\section*{Acknowledgments}
% \label{}
This research is supported by the Marie Curie TOK Action of the
European Community (project ``SPATSTAT'' MTKD-CT-2004- 014135) and
co-funded by the European Social Fund and National Resources (EPEAEK
II) PYTHAGORAS.

\end{document}